\documentclass[aps,twocolumn,prl,superscriptaddress]{revtex4}
\usepackage{amsmath}
\usepackage{graphicx}

\newcommand{\comment}[2]{\begingroup\em(#2 ---#1)\endgroup}
\DeclareMathOperator{\Tr}{Tr}

\renewcommand{\Re}{\mathop{\mathrm{Re}}}
\renewcommand{\Im}{\mathop{\mathrm{Im}}}
\newcommand{\dd}[1]{\mathop{\mathrm{d}#1}}
\newcommand{\agrad}{\hat{\nabla}}
\newcommand{\avg}[1]{\langle{#1}\rangle}
\newcommand{\gc}{\check{G}}
\newcommand{\gR}{\hat{G}^R}
\newcommand{\gA}{\hat{G}^A}
\newcommand{\gK}{\hat{G}^K}

\newcommand{\jK}{\hat{j}^K}
\newcommand{\sigmac}{\check{\sigma}}

\newcommand{\tz}{\hat{\tau}_{3}}

\renewcommand{\comment}[2]{}

\begin{document}
\title{Theory of Microwave-Assisted Supercurrent in Diffusive SNS Junctions}

\author{Pauli Virtanen$^*$}
\affiliation{%
  Low Temperature Laboratory, Aalto University School of Science and
  Technology, P.O. Box 15100, FI-00076 AALTO, Finland}

\author{Tero T. Heikkil\"a}
\affiliation{%
  Low Temperature Laboratory, Aalto University School of Science and
  Technology, P.O. Box 15100, FI-00076 AALTO, Finland}

\author{F. Sebasti\'an Bergeret}

\affiliation{Departamento de F\'{\i}sica Te\'orica de la Materia
Condensada, Universidad Aut\'onoma de Madrid, E-28049 Madrid, Spain}

\affiliation{Centro de F\'{\i}sica de Materiales (CFM), Centro Mixto
CSIC-UPV/EHU, Edificio Korta, Avenida de Tolosa 72, E-20018 San Sebasti\'an,
Spain.}

\affiliation{Donostia International Physics Center (DIPC),
Manuel de Lardizbal 4, E-20018 San Sebasti\'an, Spain.}

\author{Juan Carlos Cuevas}
\affiliation{Departamento de F\'{\i}sica Te\'orica de la Materia
Condensada, Universidad Aut\'onoma de Madrid, E-28049 Madrid, Spain}

\date{\today}

\begin{abstract}
The observation of very large microwave-enhanced critical currents
in superconductor-normal metal-superconductor (SNS) junctions at
temperatures well below the critical temperature of the electrodes
has remained without a satisfactory theoretical explanation for more
than three decades.  Here we present a theory of the supercurrent in
diffusive SNS junctions under microwave irradiation based on the
quasiclassical Green's function formalism.  We show that the
enhancement of the critical current is due to the energy
redistribution of the quasiparticles in the normal wire induced by
the electromagnetic field.  The theory provides predictions across a
wide range of temperatures, frequencies, and radiation powers, both
for the critical current and the current-phase relationship.
\end{abstract}

\maketitle

%
%
It was predicted by Eliashberg already in 1970
\cite{eliashberg1970-fss} that the condensation energy of a
superconducting thin film can be increased by irradiating the film
with microwaves. Within the framework of his theory, one can explain
the microwave-induced increase of the critical current of
superconducting bridges for temperatures very close to the critical
temperature \cite{wyatt1966-mcs,dayem1967-bot,klapwijk1977-rs},
which is known as the \emph{Dayem-Wyatt effect}. However,
Eliashberg's mechanism fails to explain a related effect in
diffusive SNS junctions. Several experiments have shown that upon
irradiation the critical current can be enhanced by up to several
orders of magnitude, even at temperatures well below the critical
temperature of the superconducting electrodes
\cite{notarys1973-jea,warlaumont1979-mpe}. Additionally, these
experiments show that the critical current is a nonlinear function
of the radiation power, which existing linear response theories
\cite{aslamazov1982-ssb,zaikin1983-nje} cannot explain.

%
%
There is now renewed interest in this problem, triggered by recent
experiments. Fuechsle \emph{et al.}\ \cite{fuechsle2009-eom}
measured the current-phase relationship under microwave irradiation,
and reported that the current is progressively suppressed at
phase differences close to $\pi$ as the radiation amplitude increases.
Moreover, Chiodi \emph{et al.}\ \cite{chiodi2009-ett} observed that
critical current is enhanced when the microwave frequency is larger
than the inverse diffusion time in the normal metal.

%
%
%
To understand the microwave-assisted supercurrent in diffusive SNS
junctions, we develop a microscopic theory based on the
quasiclassical Keldysh-Usadel approach, which takes into account the
nonlinear effects of the microwave irradiation. Our theory provides
a quantitative description for a wide range of values of the
temperature, microwave power, frequency, and the strength of
inelastic scattering. In particular, we show that the large
enhancement of the critical current originates from the presence of
a minigap, $E_g$, in the density of states of the normal wire. This
minigap blocks some of the transitions caused by the microwave
radiation, which results in a redistribution of quasiparticles,
enhancing the supercurrent when the temperature $T$ is comparable or
larger than $E_g/k_B$. We also show that the nonequilibrium
distribution in the normal wire leads to a highly
non-sinusoidal current-phase relationship, in a good agreement with
Ref.~\cite{fuechsle2009-eom}.

\begin{figure}[t]
  %
  %
  \includegraphics[width=\columnwidth]{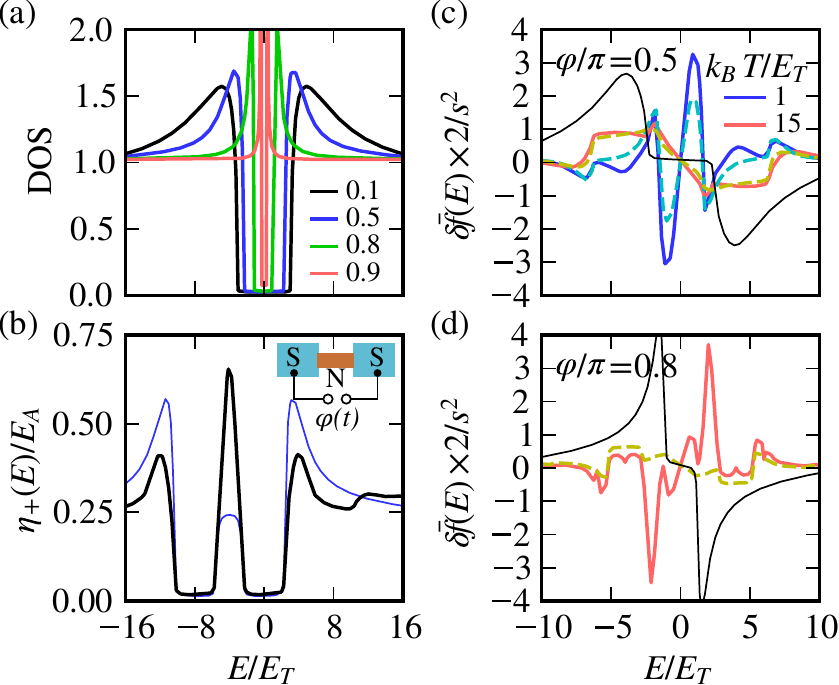}
  \caption{
    \label{fig:linearresponse} (Color online)
    (a) Local density of states (DOS) in the middle of the normal wire
    for $\Delta=100E_T$ and different values of the phase difference
    $\varphi/\pi$, in the absence of microwaves.
    (b) Absorption rate $\eta_+$ for a high frequency $\hbar\omega_0/E_T=8$
    and $\varphi=\pi/2$, $s=0.125$. Thin line shows the
    approximation from
    Eq.~\eqref{eq:eliashberg}. Inset: Schematic representation of the SNS
    junction.
    (c) Correction $\delta \bar f = \bar f-f_0$ to the electron distribution
    function vs.\ energy at two different temperatures for $\varphi=\pi/2$,
    $\hbar\omega_0/E_T=4$, and $s=0.125$. Solid lines correspond to
    the exact numerical results and the dashed lines to the approximation
    in Eq.~\eqref{eq:eliashberg}. The thin black line shows the spectral
    supercurrent $j_S(E)$ in the absence of microwaves.
    (d) The same as in (c) for $k_B/E_T = 15$ and $\varphi=0.8\pi$.
  }
\end{figure}

%
%
We consider a diffusive normal metal (N) of length $L$ connecting two
bulk superconductors with energy gap $\Delta$ (see inset of
Fig.~\ref{fig:linearresponse}(b)). In the absence of microwaves,
superconducting pair correlations leak into the normal metal modifying
its properties. For instance, the local density of states (DOS) is
modulated \cite{joyez08-stm} and a supercurrent can flow through the
normal metal \cite{dubos2001-jcc}. The DOS exhibits a minigap
$E_g(\varphi)$, see Fig.~\ref{fig:linearresponse}(a), which depends on
the superconducting phase difference $\varphi$ \cite{zhou1998-dos}.
For ideal interfaces, which we consider hereafter, $E_g(0) \approx
3.12 E_T$, where $E_T=\hbar D/L^2$ is the Thouless energy and $D$ is
the diffusion constant, whereas $E_g(\pi) = 0$.

%
%

We model the microwave radiation by an oscillating electric field,
$\vec{E}(t)$, described by a time-dependent vector potential $\vec{A}(t) =
\vec{A}_0 \cos(\omega_0 t)$, where $\vec{A}_0$ points along the axis of
the junction. We neglect screening, and assume that the
field is position independent \cite{field-footnote}.
We also neglect the effect of the radiation inside the
superconductors, which is justified for frequencies smaller than
$\Delta/\hbar$ or when the electrodes are thick compared
to the size of the junction. To evaluate the physical observables,
we use the quasiclassical theory of superconductivity for diffusive
systems~\cite{usadel1970-gde,larkin1986-ns}. It is
formulated in terms of momentum averaged Green functions
$\gc(\vec{R}, t, t^{\prime})$ which depend on position $\vec{R}$ and
two time arguments. These propagators are $4\times 4$ matrices in
Keldysh/Nambu space:
\begin{gather}
  \label{eq:keldysh-space}
  \check G =
  \begin{pmatrix}
    \hat G^R & \hat G^K \\
    0     & \hat G^A
  \end{pmatrix}
  ,\;
  \hat G^{R} =
  \begin{pmatrix}
    g^R & f^R \\
    \tilde f^R  & \tilde g^R
  \end{pmatrix}
  \,.
\end{gather}
Here, $\hat G^{R,A,K}$ are the retarded, advanced and Keldysh
components, respectively. The Green functions acquire the BCS
value $\gc_0(t-t',\pm\varphi/2)$ inside the superconductors,
and in the normal metal fulfill the Usadel equation
\begin{align}
  \label{eq:usadel}
  \hbar D \agrad\circ(\gc\circ\agrad\circ\gc) =
  [-i\epsilon\tz + i\sigmac, \gc]_\circ ,
\end{align}
where $\circ$ denotes the time convolution
$(X\circ{}Y)(t,t')=\int_{-\infty}^\infty\dd{t_1} X(t,t_1)Y(t_1,t')$,
$\agrad$ the gauge-invariant gradient $\agrad\circ X=\nabla X - i
[e\vec{A} \tz/\hbar, X]_\circ$ which involves the vector potential
$\vec{A}(t,t')=\vec{A}(t)\delta(t-t')$, and $\epsilon(t,t')=
i\hbar\partial_t\delta(t-t')$. The self-energy $\sigmac$
describes inelastic interactions in the wire, and the Green functions
are normalized as $(\gc\circ\gc)(t,t')=\delta(t-t')$.

%
%
Because Andreev reflection blocks sub-gap heat transport out of the
junction, inelastic interactions play an important role in balancing
the effect of microwaves. We describe these interactions, for
example due to phonons, within the relaxation time approximation. In
this approximation, the interaction strength is characterized by a
constant energy (scattering rate) $\Gamma$ \cite{relaxation-footnote}.
The microwave coupling, in turn, introduces the energy scale
$E_A=e^2D A^2_0/\hbar$. One can show that the ratio $s^2=E_A/E_T$
determines the change in the spectral quantities due to the microwaves,
while the ratio $\alpha = E_A/\Gamma$ controls the corresponding change
in the electron distribution. We note that $s=eV_0/\hbar \omega_0$,
where $V_0$ is the amplitude of the oscillating voltage across the junction.

%
%
In order to solve Eq.~\eqref{eq:usadel}, we follow
Ref.~\cite{cuevas2006-pea} and Fourier transform the Green functions
to energy space. Due to the time dependence of the vector potential,
the Usadel equation in energy space admits a solution of the type
$\check G({\bf R},E,E^{\prime}) = \sum_m \check G_{0,m} ({\bf R},E)
\delta(E - E^{\prime} + m \hbar \omega_0)$. With this Ansatz the
Usadel equation becomes a set of coupled differential equations for
the Fourier components $\check G_{n,m}(E) = G(E+n\hbar \omega_0,
E+m\hbar \omega_0)$. For arbitrary radiation power we solve these
equations numerically using a Jacobian-free Newton-Krylov method
\cite{knoll2003-jnm,baker2005-tac}. From the solution of $\gc$, we
can compute all physical observables.

%
%
We now analyze the linear response regime ($s^2,\alpha\ll{}1$).
In this limit, we can derive the kinetic equation for the time-average
of the distribution function, $\bar{f}$, by keeping terms up to
second order in $A_0$ in the Keldysh component of the Usadel
equation \eqref{eq:usadel}. Because of Andreev reflection, when
relaxation processes are slower than the diffusion inside the junction,
$\bar{f}$ is in our gauge constant throughout the normal wire. Consequently,
the kinetic equation reduces to an equality of the electron-phonon and
microwave collision integrals, $I_{\rm e-ph} =I_\gamma$, averaged over
the junction volume $\Omega$. The microwave collision integral resembles Joule
heating and it is proportional to a time-averaged product of electric
field and current (at energy $E$), $I_{\gamma} =
\frac{eD}{8\omega_0}\overline{\vec{E}(t) \cdot
\Tr\tz\jK(E+\hbar\omega_0/2,t)}-(E\mapsto{}E-\hbar\omega_0)$, where
$\jK = \gR\circ\agrad\circ\gK+\gK\circ\agrad\circ\gA$. Using this
result, the kinetic equation for the correction $\delta \bar{f} =
\bar{f} - f_0$, where $f_0$ is the Fermi function, becomes
\begin{align}
  \notag
  &\Gamma\avg{\rho} \delta \bar{f}
  = \eta_{-}(E+\hbar\omega_0) f_+(1-f_0) - \eta_{+}(E) f_0(1-f_+)
  \\
  \label{eq:master}
  &\quad
  + \eta_{+}(E-\hbar\omega_0) f_-(1-f_0)-\eta_{-}(E)f_0(1-f_-)
  \,.
\end{align}
Here, $\avg{\rho}$ is the spatially averaged density of states inside
the junction and $f_{\pm}=f_0(E\pm\hbar\omega_0)$. The
emission ($\eta_-$) and absorption ($\eta_+$) rates are defined
as $\eta_+(E)=\eta_-(E+\hbar\omega_0)=-\frac{eDA_0}{16} \Im\Tr\tz\avg
{\jK_{01}(E)}/(f_0(E+\hbar\omega_0)-f_0(E))$.

For frequencies $\hbar \omega_0 < 2E_g(\varphi)$, one can neglect
the AC components of the retarded/advanced functions, so that
\begin{align}
  \label{eq:eliashberg}
  \eta_+ \approx
  \frac{E_A}{4}
  \bigl\langle{
    \rho_0 \rho_+ + \Re \{(f^R_0 +
    \tilde f^{R*}_0)( \tilde f^R_+ + f^{R*}_+) \}
  }\bigr\rangle
  \,.
\end{align}
This reduces to the original linear response result by Eliashberg in
the case of a bulk superconducting film \cite{eliashberg1970-fss}.
One can now see that the minigap in $\rho$ and $f^R$ blocks some of
the radiation-induced transitions (see
Fig.~\ref{fig:linearresponse}(b)). Thus, if the temperature is
sufficiently high ($k_{\rm B}T \gtrsim E_g(\varphi)$), an excess of
quasiparticles accumulates below the minigap, and their number is
depleted above it. This cooling effect is illustrated in
Fig.~\ref{fig:linearresponse}(c), where the result of
\eqref{eq:eliashberg} is compared to the exact numerical result. As
one can see, Eq.~\eqref{eq:eliashberg} reproduces the main features
of the exact result well especially at $k_B T \gg E_g(\phi)$. Note
that despite the cooling at some energies, the Joule power
absorbed in the junction,
$P=\nu_F \int\dd{E}E \Omega \avg{I_\gamma} = \overline{I(t)V(t)}$
where $\nu_F$ is the normal state DOS, is positive.

However, Eq.~\eqref{eq:eliashberg} does not describe correctly the
behavior of the distribution function when $\hbar \omega_0>
2E_g(\varphi)$, as shown in Fig.~\ref{fig:linearresponse}(d). This
means that Eq.~\eqref{eq:eliashberg} always fails to describe the
behavior close to $\varphi = \pi$. In this limit, the radiation
induces changes in the AC components of the retarded/advanced
quantities that couple to the time-averaged distribution function,
especially at energies close to $E=\pm\hbar\omega_0/2$. Since the
behavior of these components is determined by a complicated balance
between diffusion and AC excitation, an accurate description of
$\eta_\pm$ in general requires a numerical calculation.

\begin{figure}[t]
  %
  %
  \includegraphics{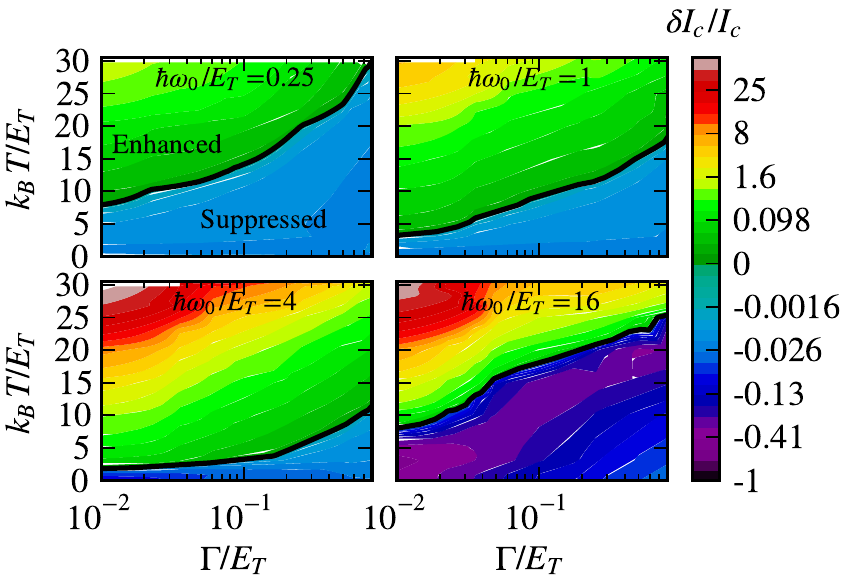}
  \caption{
    \label{fig:phasediagram}
    Correction to the critical current, normalized by the critical
    current in the absence of the field, as a function of temperature
    and inelastic rate for $\Delta=100E_T$ and several frequencies.
    The field strength is $s = 0.125$ in all cases. The lines
    separate the region of parameters for which the critical current
    is enhanced from that in which it is reduced.
  }
\end{figure}

In the limit $\Gamma \ll E_T$, the correction to the supercurrent comes
mainly from the change in the distribution function, and it can be
written as
\begin{align}
  \delta I \approx \frac{S\sigma_N}{e}\int_{-\infty}^\infty\dd{E} j_S(E)
  \delta \bar{f}(E),
\end{align}
where $S$ is the cross-section, $\sigma_N=e^2 \nu_F D$ the
normal-state conductivity of the wire and $j_S(E)$ the equilibrium
spectral supercurrent \cite{heikkila2002-sdo}, which is plotted in
Figs.~\ref{fig:linearresponse}(c) and \ref{fig:linearresponse}(d)
together with $\delta\bar{f}$. Based on this, the cooling effect
described by Eq.~\eqref{eq:eliashberg} is expected to manifest as an
enhancement of the critical current for $k_{\rm B} T \gtrsim
E_g(0)$. This is confirmed by the exact numerical calculations
obtained in the low-amplitude regime, see
Fig.~\ref{fig:phasediagram}. The effect increases up to frequency
$\hbar\omega_0 \approx {}2E_g(0)$, and at larger frequencies becomes
more varying, due to the complicated energy dependence of $j_S$ and
$\bar{f}$. On the other hand, as $\Gamma$ increases, the magnitude
of $\delta \bar{f}$ decreases, which together with the suppression
of minigap and $j_S$ reduces the current. The above is in
qualitative agreement with existing experiments
\cite{notarys1973-jea,warlaumont1979-mpe}, which concentrated on
$\hbar\omega_0/E_T\lesssim{}10$.

\begin{figure}[t]
  \includegraphics{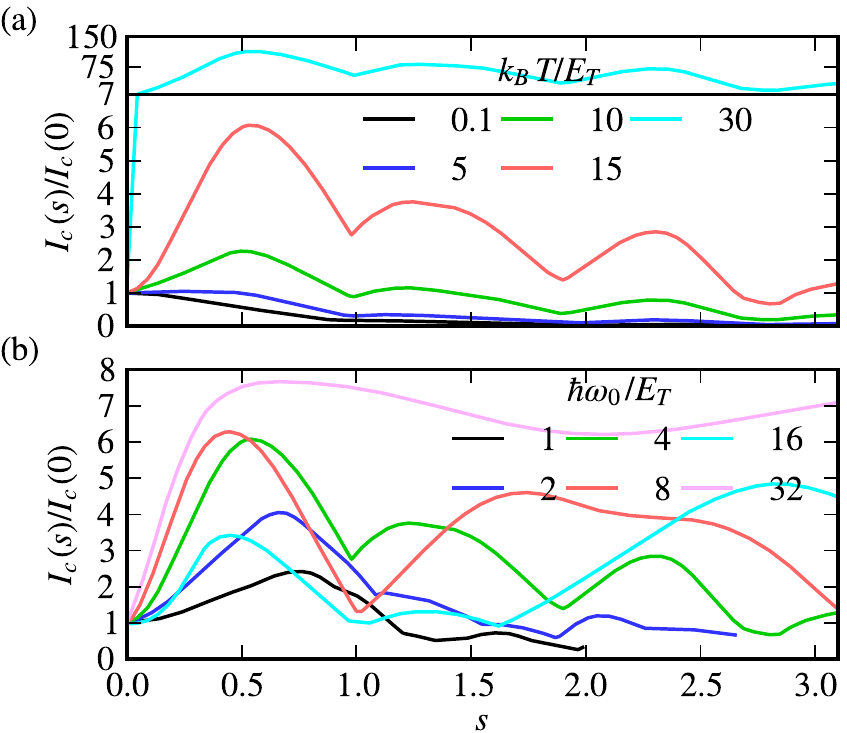}
  \caption{
    \label{fig:nonlinearresponse}
    (Color online)
    Critical current (normalized by the current without AC field)
    versus radiation amplitude $s$ for a wire with $\Delta/E_T=100$.
    (a)
    For different temperatures at $\hbar\omega_0/E_T=4$ and $\Gamma/E_T=0.05$.
    (b)
    For different frequencies and $k_BT/E_T=15$.
  }
\end{figure}

%
%

\begin{figure}[t]
  \includegraphics{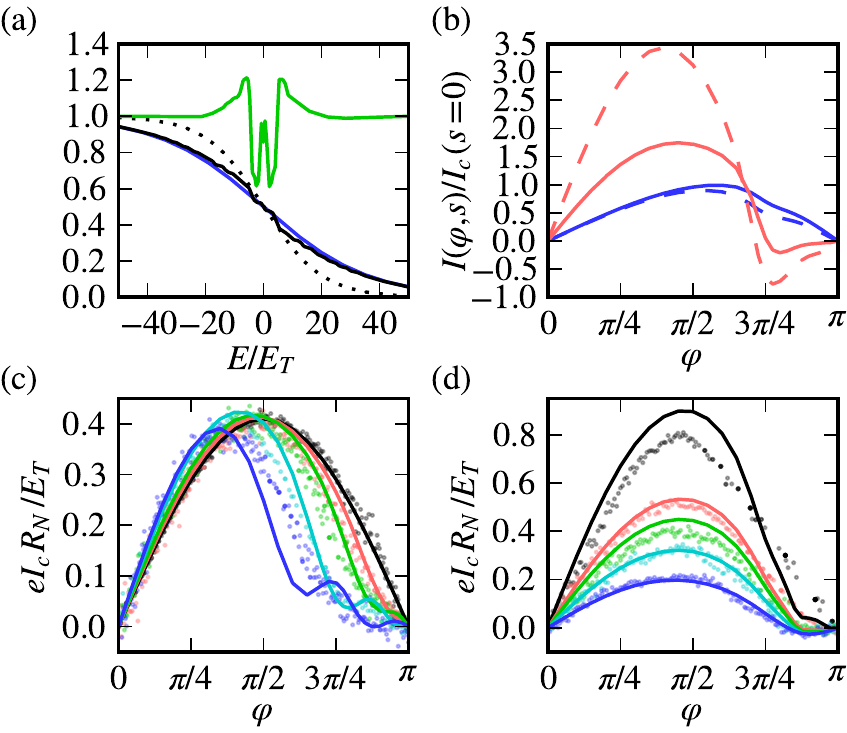}
  \caption{
    \label{fig:phasedependence}
    (a)
    Distribution function (solid black) and density of states (green)
    for large amplitude $s=2$, and $\hbar\omega_0/E_T=4$, $\Delta/E_T=100$,
    $k_BT/E_T=10$, $\Gamma/E_T=0.05$, and $\varphi=\pi/2$.
    Microwaves cause heating from $k_BT/E_T=10$ (dotted)
    to $k_BT\approx \hbar \omega_0 \sqrt{E_A/4\Gamma}$
    (blue).
    (b)
    Current-phase relation normalized to equilibrium critical current
    at $k_BT/E_T=15,1$ (top to bottom)
    and $s=0.125$ (solid) and $0.25$ (dashed), for
    $\hbar\omega_0/E_T=4$, $\Delta/E_T=100$, $\Gamma/E_T=0.05$.
    (c)
    Current-phase relation for different amplitudes $s=0, 0.2, 0.3, 0.5, 0.75$
    (solid, top to bottom) at $k_BT/E_T=10$ and $\hbar\omega_0/E_T=1.2$.
    Relaxation rate is chosen as $\Gamma/E_T=0.2 (k_BT/10E_T)^3$,
    and $\Delta/E_T=58$.
    \comment{Pauli}{
      The dip below zero near $\pi$ is difficult to reproduce with
      this large $\Gamma$.
      It may be that here details of the relaxation are important,
      as enhancement comes mainly from larger energies than suppression?
    }
    Experimental data from Ref.~\cite{fuechsle2009-eom} is shown as dots.
    (d)
    As in (c), for $s=0.3$, $\hbar\omega_0/E_T=2$,
    and temperatures $k_BT/E_T=8,9.5,10,11,12.5$ (top to bottom).
  }
  \vskip -3ex
\end{figure}

For high power, the magnitude of the critical current eventually
decreases as can be seen in Fig.~\ref{fig:nonlinearresponse}. This
occurs as large-amplitude oscillations of phase average the density
of states, which results in a suppression of the coherence and in
the subsequent closing of the minigap (see
Fig.~\ref{fig:phasedependence}(a)). As a consequence, the cooling
effect is suppressed, and microwaves mainly heat the electrons in the
same way as in the normal state, which reduces the current. In the
relaxation time approximation the temperature is for high field
strength given by $T^{\ast}\approx[P/(2\nu_F \Omega k_B^2
\Gamma_0)]^{1/5}$ 
(provided $T\lesssim{}T^{\ast}\ll{}\Delta/k_B$ and assuming $\Gamma(T)=4\Gamma_0T^3$ 
\cite{relaxation-scaling-footnote}), where $P=\sigma_N \Omega 
A_0^2\omega_0^2/2$ is the average Joule power dissipated in the 
junction.

The critical current also exhibits oscillations when radiation
amplitude increases, see Fig.~\ref{fig:nonlinearresponse}(a),
similar to those already seen in the early experiments
\cite{warlaumont1979-mpe,notarys1973-jea}. For short junctions
($\Delta < E_T$; not plotted), we find that these oscillations match
reasonably well with the usual Bessel oscillations in Josephson
tunnel junctions, i.e., $I\propto{}J_0(2s)$, but in the
long-junction limit the similarity is only qualitative. Locations of
the dips in the $I_c(s)$ relation are not strongly dependent on the
temperature, but depend on the radiation frequency, as shown in
Fig.~\ref{fig:nonlinearresponse}(b).

%

The microwave irradiation alters the current-phase relationship,
enhancing the current at $\varphi\lesssim\frac{\pi}{2}$ and
suppressing it or even changing its sign at
$\varphi\gtrsim\frac{\pi}{2}$ (see Fig.~\ref{fig:phasedependence}(b)).
The behavior near $\varphi=\pi$ comes from two sources:
the cooling disappears as the minigap closes, and the features peculiar
to the dissipative AC response of SNS junctions not contained
in Eq.~\eqref{eq:eliashberg} become increasingly important.
One can for example see in Fig.~\ref{fig:linearresponse}(d) that
at $\varphi/\pi=0.8$ the peaks at $E=\pm\hbar\omega_0/2$ give a significant
negative contribution to the current.

%
%

To compare with the results of Ref.~\cite{fuechsle2009-eom}, we
compute the current-phase relationship using the parameters of the
experiment ($T,E_T,\omega_0$). We have two free parameters:
$\Gamma/E_T$, which we assume large enough to suppress the enhancement
of the critical current, and the amplitude of the AC bias, which we
fix by assuming that $s=0.5$ corresponds to the externally applied
power level 28~dBm at $\hbar\omega_0/E_T=1.2$.  The
power dependence, see Fig.~\ref{fig:phasedependence}(c), reproduces the
main experimental features: (i) with increasing power the maximum
supercurrent is reached at $\varphi_{\rm max} < \pi/2$, (ii) the
supercurrent is strongly suppressed for phases close to $\pi$, and
(iii) for $\varphi < \varphi_{\rm max}$, the supercurrent is slightly
enhanced compared to $s=0$. On the other hand, as shown in
Fig.~\ref{fig:phasedependence}(d), the deviation from the sinusoidal
form becomes slightly more pronounced as $T$ increases, in a
qualitative agreement with experiments.  The difference to the
experiment at high power or low temperatures may be due to nonlinear
radiation coupling and the relaxation time approximation, respectively.

%
%
In summary, we have presented a general theory for describing
the effects of radiation on the properties of diffusive SNS junctions,
which explains a wide range of experimental observations. We have 
clarified the mechanism of stimulated superconductivity, shown how the
supercurrent depends on the field strength non-monotonically, and
predicted the modification of the current-phase relation. Moreover,
our results pave the way for filling some remaining gaps in the
understanding of SNS junction physics such as the finite-voltage
Shapiro steps \cite{dubos2001-jcc} or the role of phase fluctuations
providing the ``intrinsic shunting'' \cite{benz1997-s1v}.

\vskip -0.5ex

We thank Christoph Strunk, Marco Aprili, Teun Klapwijk and Yuli
Nazarov for discussions, and CSC (Espoo) for computer resources.
This work was supported by
the Spanish MICINN (contract FIS2009-04209), EC funded ULTI Project
Transnational Access in Programme FP6 (Contract RITA-CT-2003-505313).
F.S.B.\ acknowledges funding by the Ram\'on y Cajal program and T.T.H.\
the funding by the Academy of Finland and 
the ERC (Grant No. 240362-Heattronics).

%


\begin{thebibliography}{23}
\bibitem[{fie()}]{field-footnote}
\bibinfo{note}{{Electric field can be assumed constant in structures thinner
  than the skin depth, which is realistic for our parameters. We also checked
  via self-consistent calculations that the induced scalar potential
  $\phi(t)=\frac{\pi}{4}\Tr\gK(t,t)$ maintaining charge neutrality does not
  significantly affect our results.}}
\bibitem[{rel({\natexlab{a}})}]{relaxation-footnote}
\bibinfo{note}{{The simple relaxation time approximation is not exactly charge
  conserving. However, escape through the NS interfaces is the dominant charge
  relaxation mechanism, and our gauge choice implies near charge neutrality.}}
\end{thebibliography}
\end{document}